\documentclass[superscriptaddress,twocolumn,10pt]{revtex4-1}

\usepackage{amsmath,graphicx,color}
\usepackage{booktabs,array,dcolumn}

\usepackage{multirow}
\usepackage{times}
\usepackage{xcolor,soul}
\newcolumntype{C}{>{\centering}p}

\begin{document}

\title{Carbon phase diagram with empirical and machine learned interatomic potentials}

\author{George Marchant}
\email{George.Marchant@warwick.ac.uk}
\affiliation{Department of Chemistry, University of Warwick, Coventry, CV4 7AL, UK}
\author{Bora Karasulu}
\affiliation{Department of Chemistry, University of Warwick, Coventry, CV4 7AL, UK}

\author{Livia B. P\'artay}
\email{Livia.Bartok-Partay@warwick.ac.uk}
\affiliation{Department of Chemistry, University of Warwick, Coventry, CV4 7AL, UK}

\date{\today}

\begin{abstract}
In the present work we detail how the many-body potential energy landscape of interatomic potentials for carbon can be explored by utilising the nested sampling algorithm, allowing the calculation of their pressure-temperature phase diagram up to high pressures. 
We present a comparison of three interatomic potential models, Tersoff, EDIP and GAP-20, focusing on their macroscopic properties, particularly on their melting transition and on identifying thermodynamically stable solid structures up to at least 100 GPa. 
The studied models all form graphite structures upon freezing at lower pressure, then the diamond structure as the pressure increases. We were able to locate the transition between these phases in case of the Tersoff and EDIP models. 
We placed particular focus on the state-of-the-art machine learning (ML) model, GAP-20, and calculated its phase diagram up to 1~TPa to evaluate its predictive capabilities well outside of the model's fitting conditions. 
The phase diagram showed a remarkably good agreement with the experimental phase diagram up to~200 GPa, despite a variety of unexpected graphite layer spacing. Above that nested sampling identified two novel stable solid structures, a strained diamond structure and above 800 GPa a strained hexagonal-close-packed structure. However, the stability of these two phases were not confirmed by DFT calculations, highlighting potential routes to further improve the ML model.
\end{abstract}

\maketitle

\section{Introduction}

Carbon is the fourth most abundant element in the universe, and while it readily forms a much wider range of compounds than any other element (including the bio-polymers crucial for life), its behaviour is just as rich in elemental form as well. Carbon atoms can bond to each other in fascinatingly diverse ways, forming a wide range of two- and three-dimensional allotropes, amorphous phases, clusters, fullerenes and multi-layered particles that give carbon one of the most diverse ranges of chemical and physical properties among materials.\cite{C_bucky,C_electric_Novoselov,C_highP,C_nano_amorph,C_nano_book,shang_ultrahard_2021,C_onion,takagi_global_2020}
The Samara carbon database, which catalogues simulation data for these proposed structures of carbon, consists of more than five-hundred periodic configurations\cite{SACADA} (as of December 2021).
Furthermore, the properties of these structures are often unique, such as the hardness of diamond; the electronic properties of graphene; or the high ductile strength of carbon-fibres, resulting in extensive use of carbon across a wide range of industries, from battery design to advanced optical technologies.\cite{C_battery,C_optics}

One of carbon's most well known features is its phase transition from graphite to cubic diamond at pressures above 2~GPa. 
Diamond and graphite's vastly different density and structural properties are reflected in carbon's melting curve, which exhibits a dramatic change in behaviour at the corresponding triple point; shifting from a subtly non-monotonic curve at lower pressures where graphite is formed, to diamond's melting curve that quickly increases in temperature as greater pressures are applied.
Diamond remains stable up to at least 300~GPa, but due to the extreme pressure little is known experimentally of carbon's atomic structure beyond this. 
\textit{Ab initio} calculations suggest a maximum in diamond's melting line at around 450~GPa, as well as a transition to bc8 between 890-1000~GPa, and shock-wave experiments provide evidence for the accuracy of these predictions.\cite{C_highpressure_exp,C_pressure_rev}  The bc8 structure is also predicted to have a maximum in the melting temperature of around 1450~GPa, due to changes in the coordination number in the liquid phase.\cite{C_DFT_highP_diagram} 
In the terapascal regime further phase transitions are predicted, such as bc8-simple cubic and simple cubic-simple hexagonal.\cite{C_highP} 

Atomistic simulations have thus played a major role in discovering novel phases of carbon; furthering our understanding of its phase diagram; and driving the development of new applications by providing useful insight into their structure and properties.
However, the diverse properties of carbon mean that capturing its various characteristics within interatomic potential models is particularly difficult, especially when creating models that aim to be transferable among different allotropes, and reproduce carbon's macroscopic properties reliably under a wide range of conditions. 

Several empirical interatomic potential models have been developed for carbon in the past 35 years.
The many-body potential introduced by Tersoff\cite{tersoff_potential} in 1988 is still considered to be the fastest and most simple carbon potential. 
Its elegant functional form, in which the strengths of chemical bonds are modified according to the number of nearest neighbours, allows for rapid calculation of chemical properties without a significant sacrifice in accuracy when compared to other, more expensive potentials.\cite{carbon-potential-review} 
Despite its shortcomings - the primary one being its lack of consideration for long-range interactions - it is still an ideal choice for testing the performance of new computational methods and more complex chemical potentials. 
Other early carbon models include the Stillinger-Webber potentials parameterised for diamond and graphitic carbon,\cite{SW_diamond,SW_graphite} although, due to using fixed coordination, these models are limited in their transferability across structures.
Developed from the Tersoff model to include a wider range of parameters, conjugation and torsional terms, the reactive bond order potentials were introduced: REBO (also referred to as the Brenner potential)\cite{brenner} and REBO-II\cite{REBO-II}.
These bond order potentials were further improved by the inclusion of a long-range term to create a potential that accounts for the effects of dispersion, providing the adaptive intermolecular REBO (AIREBO).\cite{AIREBO}
The environment-dependent interaction potential (EDIP) consists of a two-body pair energy, a three-body angular penalty, as well as a generalized description of coordination.\cite{marks_edip}
EDIP is known to successfully predict topological properties of carbonecous films as well as clusters.\cite{carbon-potential-review_2019,karasulu_C_2022}
One of the first empirical models capable of providing an accurate description of low to medium pressure phases of carbon is the long-range carbon bond-order potential (LCBOP). The LCBOP model is partially based on \textit{ab initio} data, closely matches the \textit{ab initio} MD results for the liquid structure, and accounts for interplanar interactions in graphite.\cite{LCBOB_2003}
Ghiringelli et al. have calculated the pressure-temperature phase diagram of the LCBOP potential, calculating the melting line up to 60~GPa and graphite-diamond transition, showing a good agreement with experimental findings.\cite{ghiringhelli_LCBOP_PD} 
Another family of potentials were developed to accurately describe carbon's bond formation and dissociation: the reactive force field (ReaxFF) potentials.\cite{reaxff,reaxff_2013} 

The emergence of machine-learning techniques offer the construction of potential models which are comparable in cost to classical interatomic potentials, but comparable in accuracy to ab initio level calculations. Using the Gaussian Approximation Potential (GAP) formalism\cite{bartok_2010,bartok_2013} a machine-learned potential was developed to describe the behaviour of liquid and amorphous carbon accurately.\cite{GAP_2017}
This was later extended to include properties of crystalline bulk phases, defects and surfaces, known as the GAP-20 model.\cite{GAP_20,GAP20_erratum}
The C60 GAP force field includes van der Waals corrections and is especially suited for the simulation of  C$_{60}$ fullerene structures.\cite{muhli_machine_2021}
Recently, another ML carbon potential was developed as well, using neural-networks. \cite{carbon_NNP}

The performance and reliability of these potentials have been compared from different perspectives. 
Their ability to describe amorphous structures\cite{carbon-potential-review,carbon-potential-review_2019} has demonstrated a lack of transferability and highlighted the need for thorough investigation of models in order to trust the interpretation of simulation results.
The accuracy in predicting microscopic properties (e.g. surface energy, formation energy of common defects) have been also compared.\cite{GAP_20}
The performance of seven models in predicting the properties of carbon nano-clusters have been recently investigated, with a focus on their accuracy in structure search and global optimisations.\cite{karasulu_C_2022} The GAP-20 model has emerged as the best performing model.

While these studies provide a detailed picture on the microscopic properties of carbon potentials, our knowledge of their macroscopic properties is limited. In order to understand the reliability and predictive power of computational results, it is important to examine the potential models' macroscopic behaviour and evaluate their phase stability, unbiased by our chemical intuition. 
Ultimately, this also informs the development of new generations of potentials, such as machine learning-based models, highlighting strengths as well as areas for improvement.

In the current work we aim to extend our knowledge of the performance of carbon models and calculate their pressure-temperature phase diagram, by performing an exhaustive and predictive sampling of the entire potential energy surface, using the nested sampling technique.\cite{NS_mat_review,NS_all_review}
Nested sampling (NS) was first introduced by John Skilling in the area of Bayesian statistics, \cite{bib:skilling,bib:skilling2} later taken up by various research fields\cite{NS_all_review} and adapted to sample the potential energy surface of atomistic systems\cite{1st_NS_paper, NS_mat_review}. 
The main advantages of NS are that it automatically generates the thermodynamically relevant structures without any prior knowledge of e.g. crystalline structures, and it provides unique and easy access to the notoriously elusive partition function. Thermodynamic properties that are otherwise difficult to determine, such as the heat capacity or free energy, thus become straightforwardly calculable.
The power of NS has been demonstrated in studying various systems, including the formation of clusters,\cite{1st_NS_paper,CuPt_ns,dorrell_thermodynamics_2019} calculation of the quantum partition function,\cite{szekeres_direct_2018} sampling transitions paths,\cite{bolhuis_nested_2018} as well as the calculation of the pressure-temperature phase diagram for various metals,\cite{pt_phase_dias_ns,ConPresNS,NS_lithium} alloys,\cite{CuAu_Pastewka,AgPd_ML} and model potentials,\cite{jagla} identifying previously unknown stable solid phases.
Using NS, we revisit two widely used classical many-body potentials, but our main focus is the Gaussian Approximation Potential developed by Rowe et al\cite{GAP_20} and its updated version\cite{GAP20_erratum}, 
to examine its reliability outside of the original training conditions and hence understand the extent of the model's transferability and predictive power.


\section{Computational details}

\subsection{Studied potential models}

In the current work we compare the behaviour of three 
widely used interatomic potential models for carbon, spanning a range in complexity, accuracy and computational cost. 
We first use the machine-learning potential, GAP-20 \cite{GAP_20}, considered to be the state-of-the-art model for carbon.\cite{C_highP,carbon-potential-review,carbon-potential-review_2019} The majority of our GAP-20 calculations were performed using the original model detailed in Ref.~\cite{GAP_20}, and we also provide supplemental results generated with the updated version of the model, GAP-20u, released very recently. 
As the fastest and simplest model, we evaluate the phase diagram of the Tersoff model in the original parameterisation form, as available in LAMMPS (although valuable modifications to the Tersoff carbon potential also exist).
We also selected the environment-dependent interaction potential (EDIP) \cite{marks_edip} for modelling, providing a mid-point in accuracy and computation cost between the Tersoff and GAP-20 potentials.


\subsection{Nested sampling}

The nested sampling calculations were performed as presented in~\cite{pt_phase_dias_ns}. 
After the sampling has been done, we calculate the partition function and derived thermodynamic response functions to determine the phase behaviour. We use the position of peaks in the heat capacity to locate phase transitions, and calculate the phase space-weighted averages of observables (e.g. coordination number) to evaluate their finite temperature values using the following equation:
\begin{equation}
    \langle A \rangle \approx \frac{1}{\Delta}\sum_i A_i (\Gamma_{i-1}-\Gamma_{i}) e^{-\beta H_i},    
\label{eq:average}
\end{equation}
where $\Delta$ is the isobaric partition function; $\beta$ is the inverse temperature; $A_i$, $H_i$ and $\Gamma_i$ are the observable value, enthalpy and phase space volume of the $i$-th configuration respectively, where $\Gamma_i=(K/K+1)^i$ and $K$ is the number of walkers in the simulation.

In an infinite system the heat capacity peaks would be divergent due to discontinuity of the first order in the corresponding enthalpy vs. temperature curves, but the finite size of these systems cause a broadening of the peaks. The temperature of a given transition and its error are ascertained from the combination of data from each of the three independent runs we performed at every pressure in order to test the convergence of the simulations: fits to Gaussian functions are performed in the region of each peak and the lower and upper bounds of the error are taken to be the minimum and maximum temperature values of the peaks' half-maximums. 
The simulations were run at constant pressure, and the bounding cell of variable shape and size contained 16, 32 or 64 particles, to be able to estimate the finite size effect. Previous calculations show that the small system size usually causes the melting temperature to be overestimated, however, the solid-solid transitions are less affected, with sampled crystalline phases usually remaining consistent across different system-sizes.
For each calculation, the number of walkers, $K$, were chosen such that the resulting heat capacity peaks were sufficiently converged, thus predicted transition temperatures were generally within a range of 200K (exceptions are noted). Using a larger number of walkers means a sampling of higher resolution, with the computational cost increasing linearly with $K$. 
Initial sample configurations were generated randomly, with new samples generated by performing Hamiltonian Monte Carlo (all-atom) moves, as well as changing the volume and shape of the simulation cell by shear and stretch moves, with a step probability ratio of 5:3:2:2, respectively. 

A parallel implementation of the NS algorithm is available in the {\tt pymatnest}  python software
package~\cite{pymatnest}, using the LAMMPS package~\cite{LAMMPS} for the dynamics.

\begin{table}[]
    \centering
    \begin{tabular}{ |p{2.2cm}|c|c|c|c|c|c|c| } 
    \hline
    \multirow{2}{*}{\textbf{NS parameters}} &  \multicolumn{2}{c|}{\textbf{GAP-20}} & \multicolumn{2}{c|}{\textbf{EDIP}} & \multicolumn{3}{c|}{\textbf{Tersoff}} \\
    \cline{2-8}
    \rule{0pt}{0.3cm} & 16 atoms & 32 & 16 & 32 & 16 & 32 & 64\\
    \hline
    $K$ & 640 & 528 & 1050 & 1470 & 1680 & 2520 & 2880 \\
    \hline
    \# Model calls & 640 & 500 & 840 & 840 & 1120 & 1260 & 1440 \\
    \hline
    \end{tabular}
    \caption{Summary of nested sampling parameters used in the exploration of the potential energy surface of the three different studied carbon models. The number of walkers, $K$, determine the resolution of the sampling, the number of models calls is the average number of energy evaluations per NS iteration, while generating a new sample.}
    \label{tab:my_label}
\end{table}

\subsection{DFT calculations}

In order to compare the energies of configurations and phase stability predicted by the GAP-20 and GAP-20u models, we employ Density Functional Theory (DFT) with the same input parameters as those used to generate the data on which the GAP-20u model was trained.
DFT calculations are therefore carried out using the Vienna \textit{ab initio} Simulation Program (VASP), with the dispersion-inclusive optB88-vdW exchange-correlation functional,\cite{DFT_XC_vdW_1, DFT_XC_vdW_2, DFT_XC_vdW_3, DFT_XC_vdW_4} and the Projector Augmented Wave (PAW) pseudopotential method (PAW\_PBE C 08Apr2002)\cite{DFT_plane-wave_1, DFT_plane-wave_2, DFT_plane-wave_3} with a plane-wave cutoff of 600 eV. In each case, reciprocal space is sampled using an automatically generated Monkhorst-Pack mesh such that the smallest spacing between k-points is no greater than 0.2~\AA$^{-1}$, and energy levels are smeared by Gaussian distributions with widths of 0.1 eV.

\section{Results}

\subsection{GAP-20}
\begin{figure}[hbt]
\begin{center}
\includegraphics[width=8.8cm,angle=0]{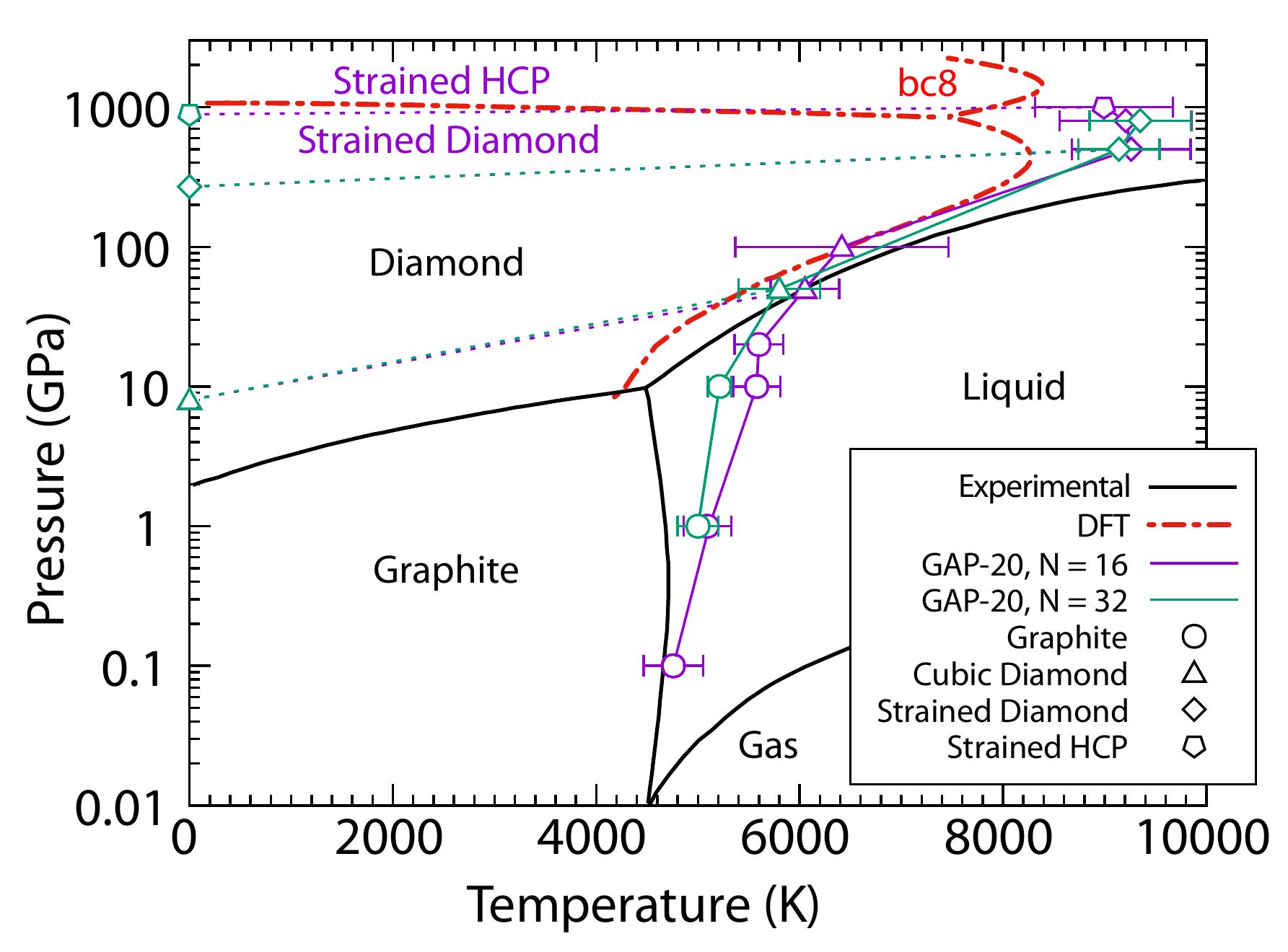}
\end{center}
\vspace{-20pt}
\caption {Pressure-temperature phase diagram of the GAP-20 potential. Black lines show experimental phase boundaries\cite{bundy1989pressure,bundy1996pressure, steinbeck1985model}, red dashed lines show high-pressure phase transitions predicted by DFT from Ref.\cite{C_DFT_highP_diagram} (with the phase above 1~TPa being bc8), purple and green lines and symbols show nested sampling results with different system sizes. Symbols reflect the most stable phase predicted by nested sampling at the corresponding temperature and pressure. Error bars represent the full widths at half maximum of the heat capacity peaks.}
\label{fig:GAP_phase-diag}
\end{figure}

Nested sampling runs with the GAP-20 potential were carried out with a system size of 16 atoms at ten different pressures between 0.1 and 1000 GPa. 
Due to the large computational cost of the GAP potential, fewer calculations were carried out with 32 atoms - at pressures of 1, 10, 50, 500 and 800 GPa - to assess the finite size effects at pressures where different solid phases are expected. 
The resulting pressure-temperature phase diagram is illustrated in Figure~\ref{fig:GAP_phase-diag}.
The experimentally determined phase boundaries \cite{bundy1989pressure,bundy1996pressure, steinbeck1985model} are shown by solid black lines, highlighting that the graphite melting line has a  slight maximum, as above 0.4~GPa the density of graphite becomes lower than that of the liquid, causing the melting line to have a negative gradient. This change however is very subtle, driven by the relatively weak interaction between graphite's neighbouring hexagonal layers.  
Above 20~GPa the liquid carbon freezes into the high-density cubic diamond structure, resulting in melting temperatures increasing rapidly with pressure in comparison to the graphite phase. 

\begin{figure}[hbt]
\begin{center}
\includegraphics[width=8.5cm,angle=0]{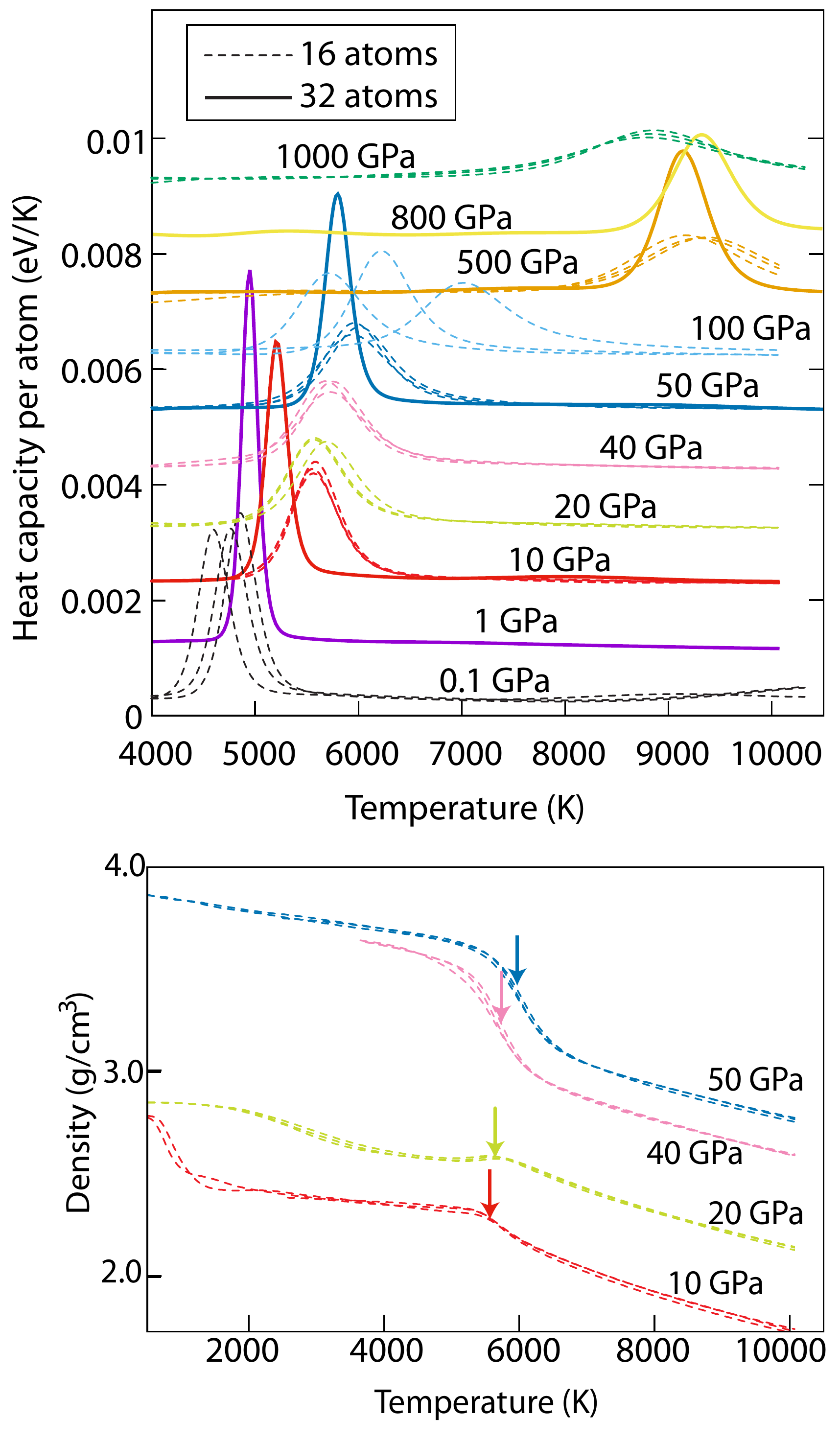}
\end{center}
\caption {Top panel: heat capacity as a function of temperature at various pressures, calculated using the GAP-20 with 16 atoms (dashed lines) and 32 atoms (solid lines). Heat capacities are shifted vertically for better visibility.
Bottom panel: density as a function of temperature, calculated using the GAP-20 potential, using 16 atoms. Arrows point to the temperatures at which the peaks of the corresponding heat capacity curves are found.
}
\label{fig:GAP_hc_all}
\end{figure}

\begin{figure}[hbt]
\begin{center}
\includegraphics[width=8.5cm,angle=0]{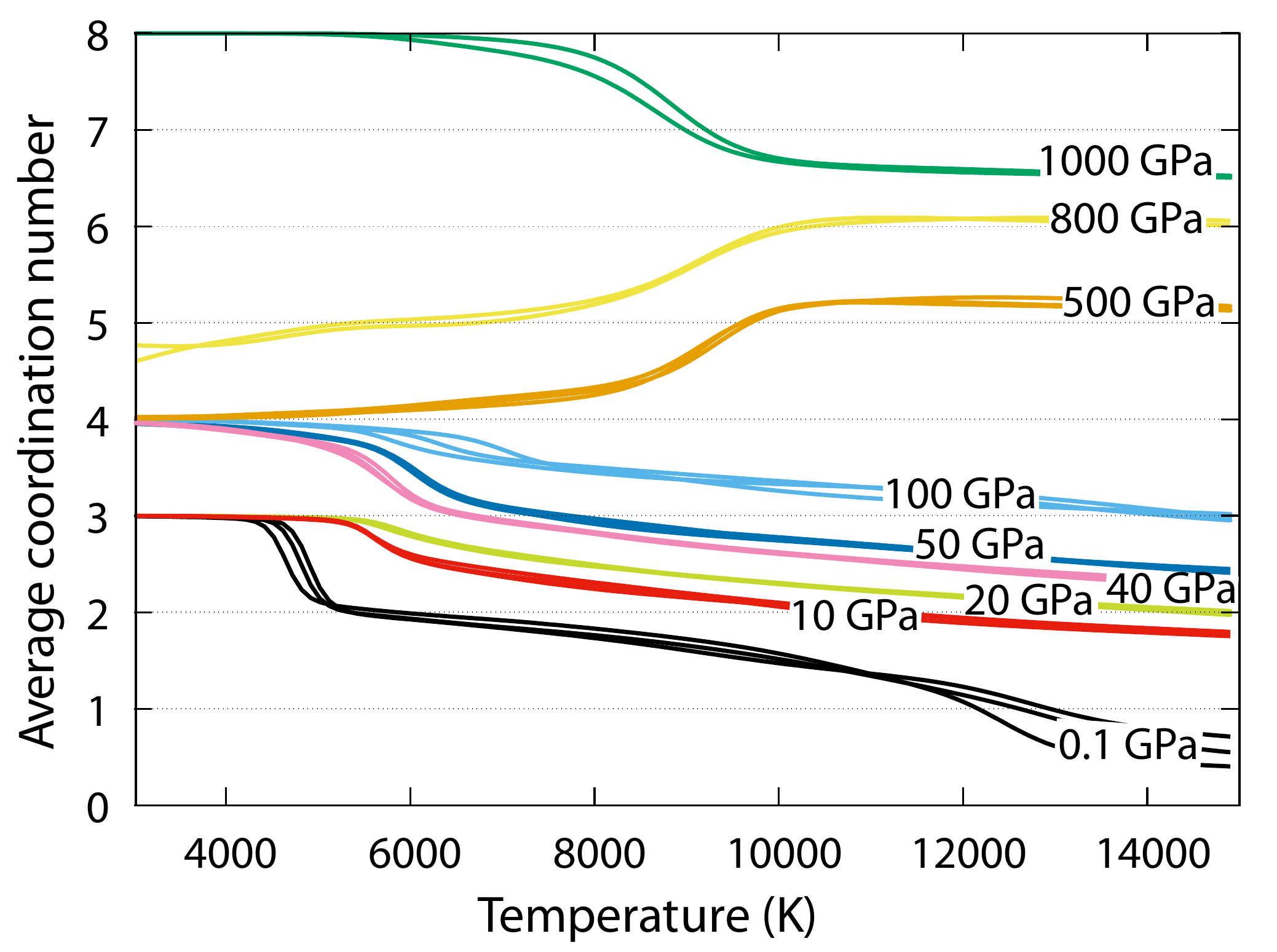}
\end{center}
\caption {Average coordination number, within a cutoff of 1.8~\AA~as a function of temperature using GAP-20 with 16 atoms, sampled by NS. Multiple lines correspond to results of multiple parallel NS runs.}
\label{fig:GAP_coordination}
\end{figure}

The melting curve predicted by the GAP-20 model follows these experimental features closely, with the melting temperature not changing significantly between 0.1 and 20 GPa.  
~Figure\ref{fig:GAP_hc_all} shows the heat capacity curves calculated by nested sampling, using the GAP-20 model. 
Lines corresponding to 32-atom runs are sharper than that of 16-atom runs, reflecting that at the thermodynamic limit the heat capacity diverges at first order phase transitions. 
The difference between the transition temperature predicted by 16 and 32-atom runs is approximately 8\% at lower pressures, with the difference diminishing at pressures above 100 GPa, suggesting that finite size effects become negligible at higher pressures. 
At 0.1~GPa, the liquid phase generated by NS is dominated by chain-like structures. 
This is in agreement with the known low-coordinated liquid phase formed at low pressures, dominated by branch-like structures.\cite{GAP_2017} 
To demonstrate the change in the typical coordination of carbon atoms at different temperatures and pressures, we calculated the NS weighted average of the coordination number over a range of temperatures using Eq.~\ref{eq:average}, shown in Figure~\ref{fig:GAP_coordination}. 
This figure shows that the at 0.1 GPa the average number of neighbours in the liquid phase is two, which increases to around three at 10 GPa, four at 100 GPa and eight at 1000 GPa. We also note that, as expected, the coordination numbers converge upon a value of three at pressures where graphite is formed, and four in the case of diamond.

Up to 20 GPa the liquid freezes into the graphite structure.
While at lower pressures the density of the graphite is found to be higher than that of the liquid, this trend changes, and at 20~GPa we can observe a maximum on the density curve, see bottom panel of Figure~\ref{fig:GAP_hc_all}, suggesting that the melting line has a negative gradient in that range. 

\begin{figure}[hbt]
\begin{center}
\includegraphics[width=9cm]{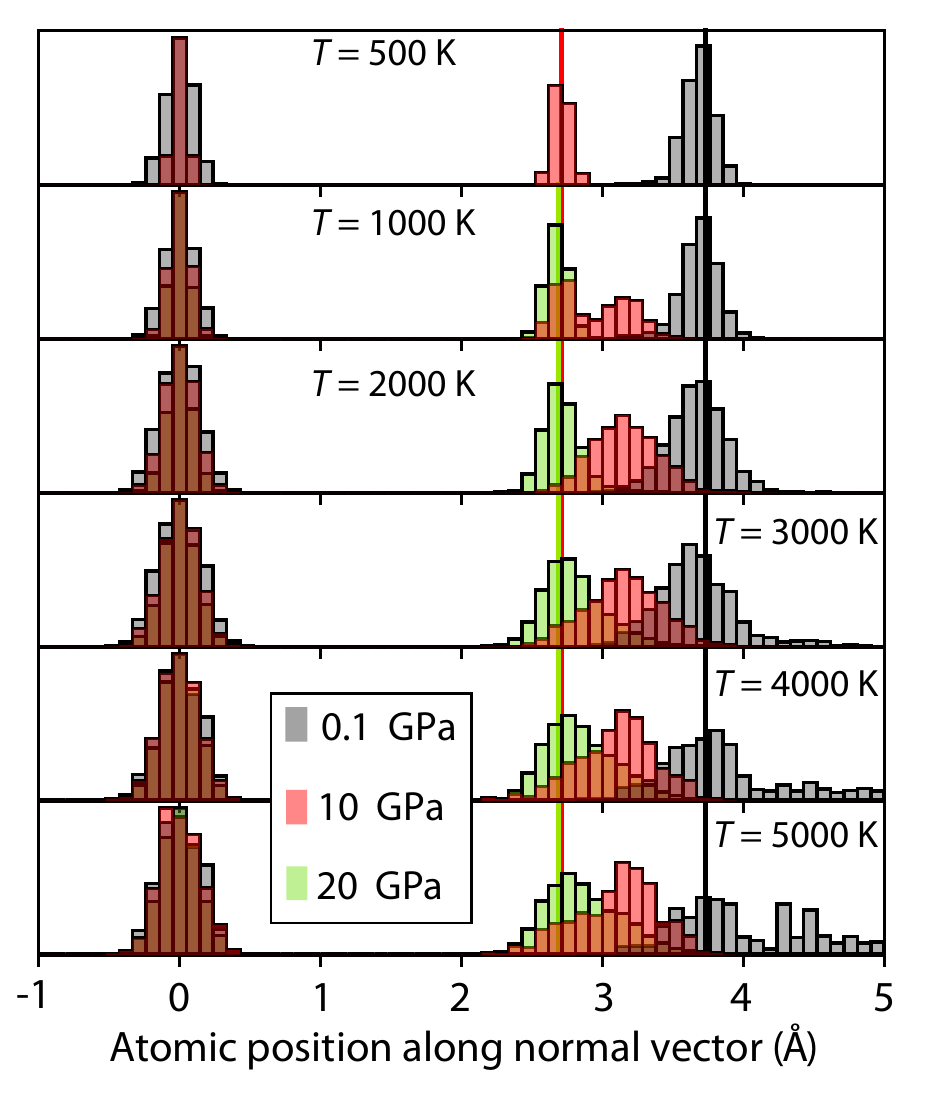}
\end{center}
\vspace{-20pt}
\caption {Distribution of the carbon atoms along the direction perpendicular to the graphite layers, calculated as the weighted average from the NS configurations, using the GAP-20 potential and 16-atom runs. Distribution around 0.0~\AA~acts as the reference layer, its widths representing the deviation from a perfectly flat layer. }
\label{fig:layers}
\end{figure}

\begin{figure}[hbt]
\begin{center}
\includegraphics[width=8.5cm,angle=0]{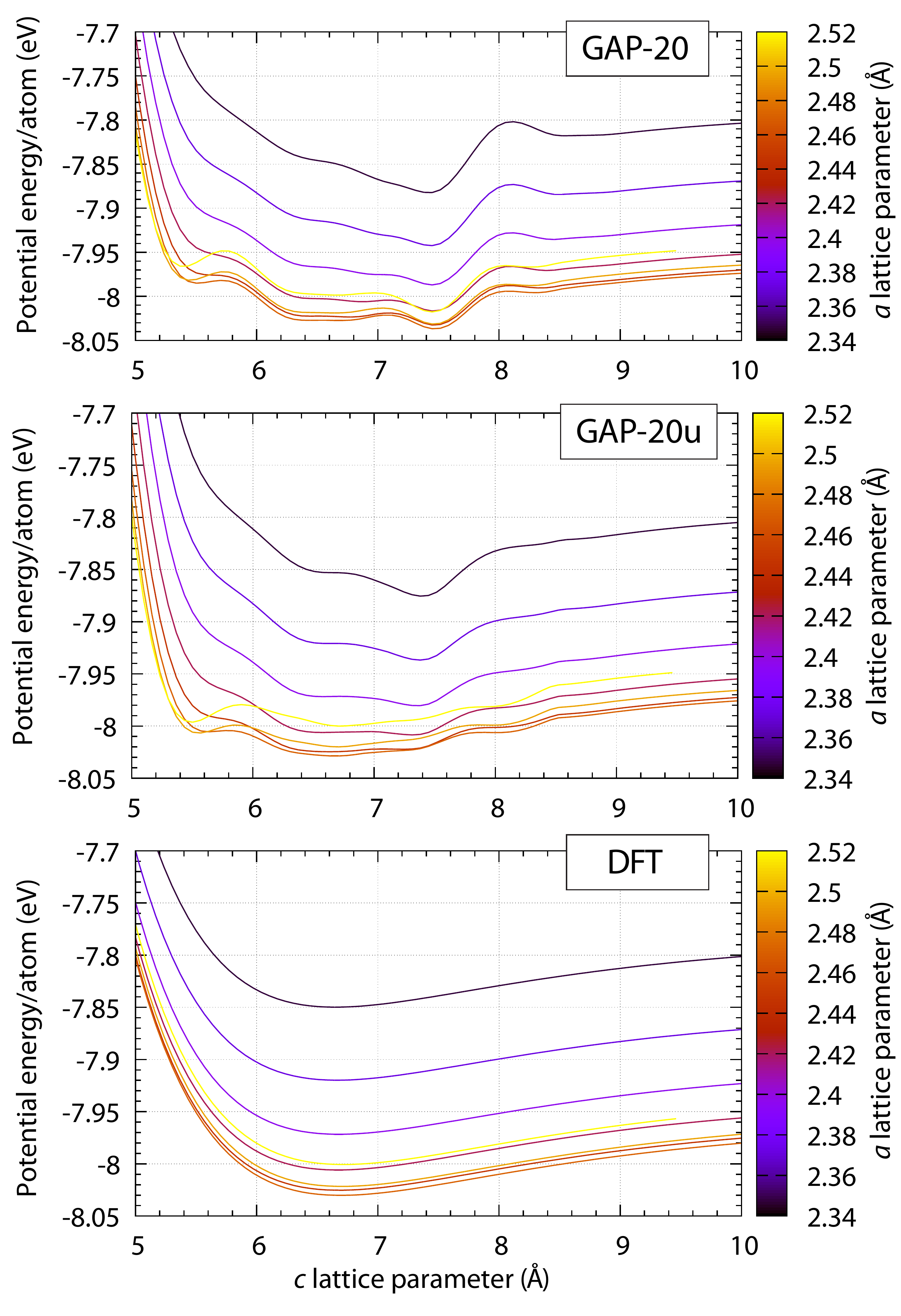}
\end{center}
\caption {Potential energy of the graphite AB structure as a function of lattice parameters, calculated using GAP-20 (top panel), GAP-20u (middle panel) and DFT (bottom panel).}
\label{fig:GAP_graphite-lattice}
\end{figure}

The explored graphite configurations are rich both in the distance of the graphite layers as well as in the relative orientation  stacking patterns.
Among the configurations generated by NS, we can find the energetically most favourable AB stacking variant,\cite{graphite_cancado_2008,graphite_wigner_2003} as well as ABC and a range of different stackings, with layers shifted only partially in relation to each other, often in multiple directions. 
In terms of the distance between the graphite layers, we see a significant change with respect to temperature and pressure. 
Figure~\ref{fig:layers} shows the distribution of carbon atoms along the normal vector of the graphite structure at different pressures and temperatures, calculated as the phase space weighted average (using eq.\ref{eq:average}) from configurations generated by NS.
At 0.1~GPa the typical spacing between neighbouring layers is around 3.8~\AA, with the initially broad distribution getting significantly narrower as the temperature decreases. This layer distance corresponds to lattice parameter $c=7.6$~\AA~in the AB stacking, much larger than the experimentally observed $c=6.7$~\AA.
The underlying reason to this discrepancy becomes obvious by calculating the energy of graphite structures as a function of the lattice parameters, shown in Figure~\ref{fig:GAP_graphite-lattice}. 
These calculations reveal that the graphitic energy surface has multiple minima with respect to layer spacing in the case of GAP-20, with the lowest energy distance confirmed to be at $c=7.6$~\AA.
At higher pressures the contribution of the pressure-volume term grows in comparison to the potential energy, and as such minima corresponding to shorter layer distances become favourable in terms of enthalpy. This is reflected in the histograms of Figure~\ref{fig:layers}, showing that NS at 10 and 20 GPa sampled closer graphite layering. We even observe a phase transition at 10 GPa between $500$ and $2000$~K, as the average spacing quickly increases from $c=5.5$~\AA ~to $6.3$~\AA, with a double peak feature at $1000$ K reflecting the simultaneous sampling of graphite basins with distinct layer spacings.
It has to be noted that this behaviour naturally affects the density of the sampled graphite phases as well, with low pressure graphite being less dense than expected and more dense at higher pressures. The densities in Figure~\ref{fig:GAP_hc_all} demonstrate this, as we see a significant difference in the density of the solid phases at 10 and 20 GPa, as well as a sudden jump in density at 10 GPa corresponding to the previously noted graphitic phase transition around $1000$~K. We speculate that this behaviour, if affecting the density ratio between graphite and liquid carbon, is capable of notably changing the gradient of the melting curve and shifting the expected maximum in the melting temperature to higher pressures.
The multiple minima as a function of graphite lattice parameters can be still observed, though to a lesser extent, in the case of GAP-20u (see Figure~\ref{fig:GAP_graphite-lattice} middle panel), though DFT reference calculations (shown in bottom panel) suggest that these curves should be completely smooth, with only a single minimum at $6.7$~\AA.

Although nested sampling simulations at 40, 50 and 100~GPa 
also explored graphite and hexagonal diamond structures to some extent, these phases remain metastable at any temperatures, as the cubic diamond structure becomes the dominant phase.
Crucially, the change in the stable solid phase, from graphite to diamond, also corresponds to the change from a roughly vertical curve to one with a large positive gradient, as also observed experimentally.\cite{bundy1989pressure,bundy1996pressure, steinbeck1985model}
These agreements are particularly notable, as the GAP-20 model was only trained on 0~GPa configurations, with high-pressure behaviour not considered in the potential development process. 

\begin{figure}[hbt]
\begin{center}
\includegraphics[width=8cm]{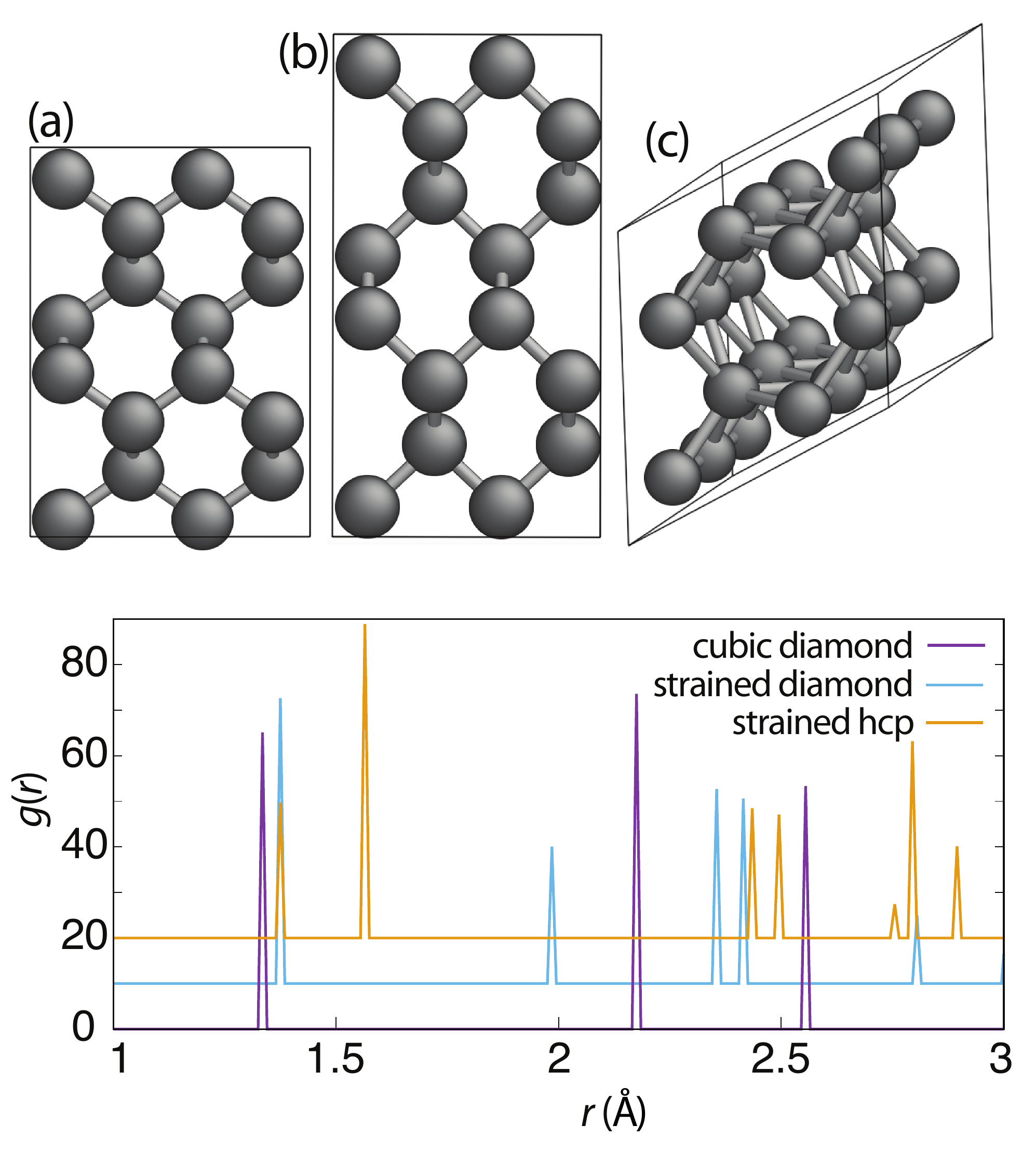}
\end{center}
\vspace{-20pt}
\caption {High pressure structures found to be stable by nested sampling, using the GAP-20 potential. Cubic diamond (a) strained diamond (b) and strained hexagonal-close-packed (c) structures. Lower panel shows the radial distribution function of the above three structures, optimised at 300~GPa.}
\label{fig:strained}
\end{figure}

\begin{figure}[hbt]
\begin{center}
\includegraphics[width=8.5cm,angle=0]{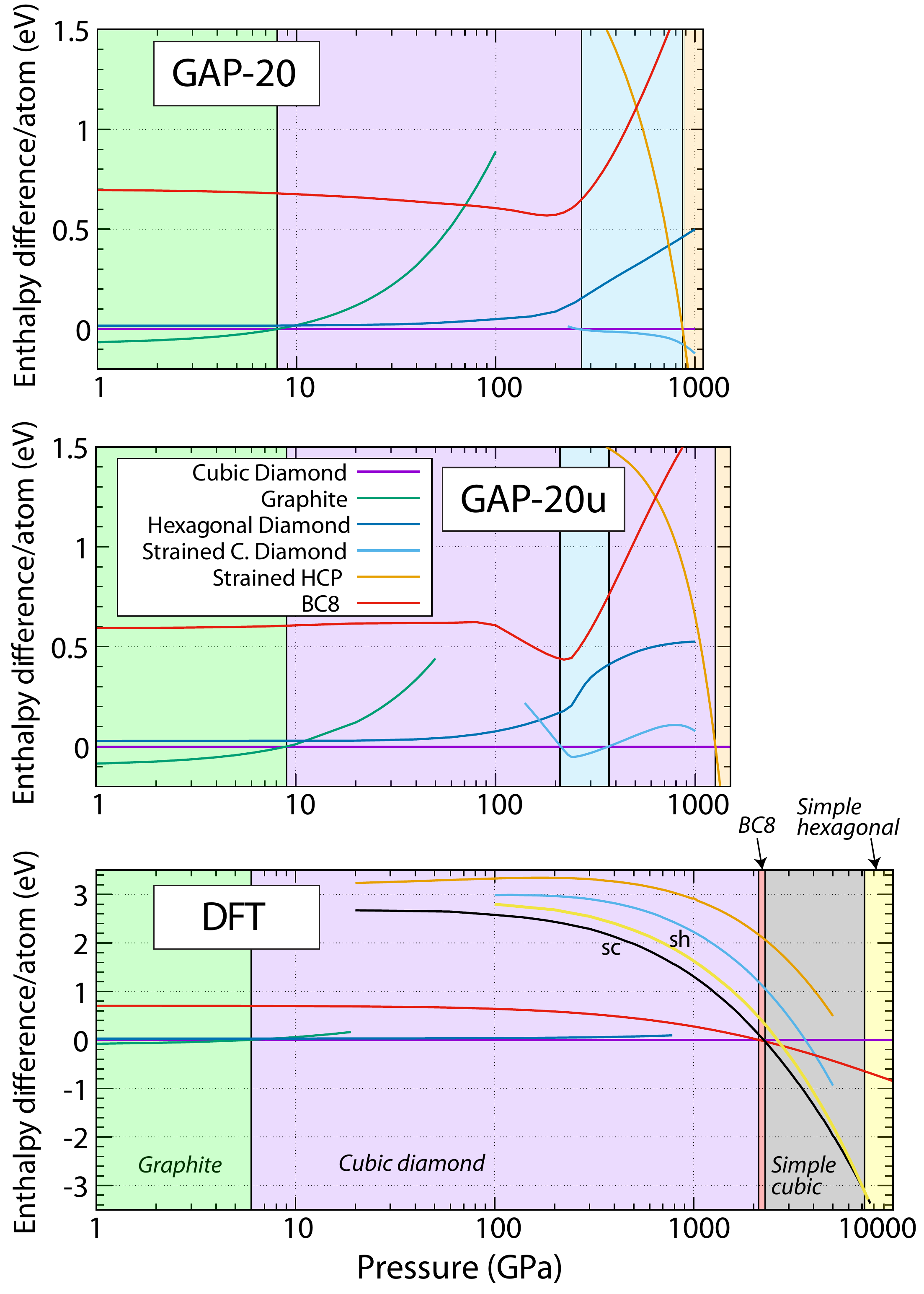}
\end{center}
\caption {Enthalpy of several thermodynamically relevant carbon crystal structures, calculated as a function of pressure using the GAP-20 potential (top), GAP-20u potential (middle), and DFT 
(bottom). In each case, energies are compared to the cubic diamond structure, and shaded areas highlight the pressure region when a structure is the ground state. Simple cubic structure is marked with \textit{sc}, simple hexagonal with \textit{sh} in the bottom panel.}
\label{fig:GAP_HvP}
\end{figure}

The predictive power of the model weakens eventually, and at extreme high pressure two new phases emerge as ground state structures of the model, both in the 16-atom and 32-atom simulations.
At 500 and 800~GPa the stable structure predicted by nested sampling is that of a strained variant of cubic diamond, where the strain is positive, in the direction of an arbitrary cubic axis and coupled with a compression along the perpendicular axes. 
Between 800 and 1000 GPa the system transitions to a highly compressed hexagonal close packed structure.
This belongs to the P6$_3$/mmc spacegroup, having two atoms in the unit cell, each with eight nearest neighbours. We will refer to this structure as strained hexagonal close-packed (strained hcp).
Figure~\ref{fig:strained} shows snapshots of these two new structures along with the cubic diamond, as well as the corresponding radial distribution functions, with all three structures optimised at 300~GPa. 
The enthalpy difference between the different optimised structures at 0~K are shown in Figure~\ref{fig:GAP_HvP}, calculated by the GAP-20 and GAP-20u potential up to 1 TPa, as well as with DFT for comparison up t0 10 TPa.
While both GAP-20 and GAP-20u predicts the appearance of the strained cubic diamond structure at very high pressures, cubic  diamond becomes the ground state again above 380 GPa in case of GAP-20u. This shows, unsurprisingly, that the changes caused by the refitting affected the behaviour of the model much more far away from the fitting conditions. 
One should also note that as the bc8 structure was not included in the training data, neither versions of the GAP model could predict it to be a low-enthalpy state at pressures above 1 TPa.
Geometry optimisations using the same DFT parameters as used in the training, show good agreement with previous \textit{ab initio} random structure search results\cite{C_highP}, showing a ground state transition from cubic diamond to bc8, simple cubic (sc), then to simple hexagonal (sh). 
While neither of the high-pressure configurations predicted by the GAP model proves to be ground state structures, they are nevertheless low-enthalpy states worth considering.

\subsection{Tersoff potential}

\begin{figure}[hbt]
\begin{center}
\includegraphics[width=8.5cm,angle=0]{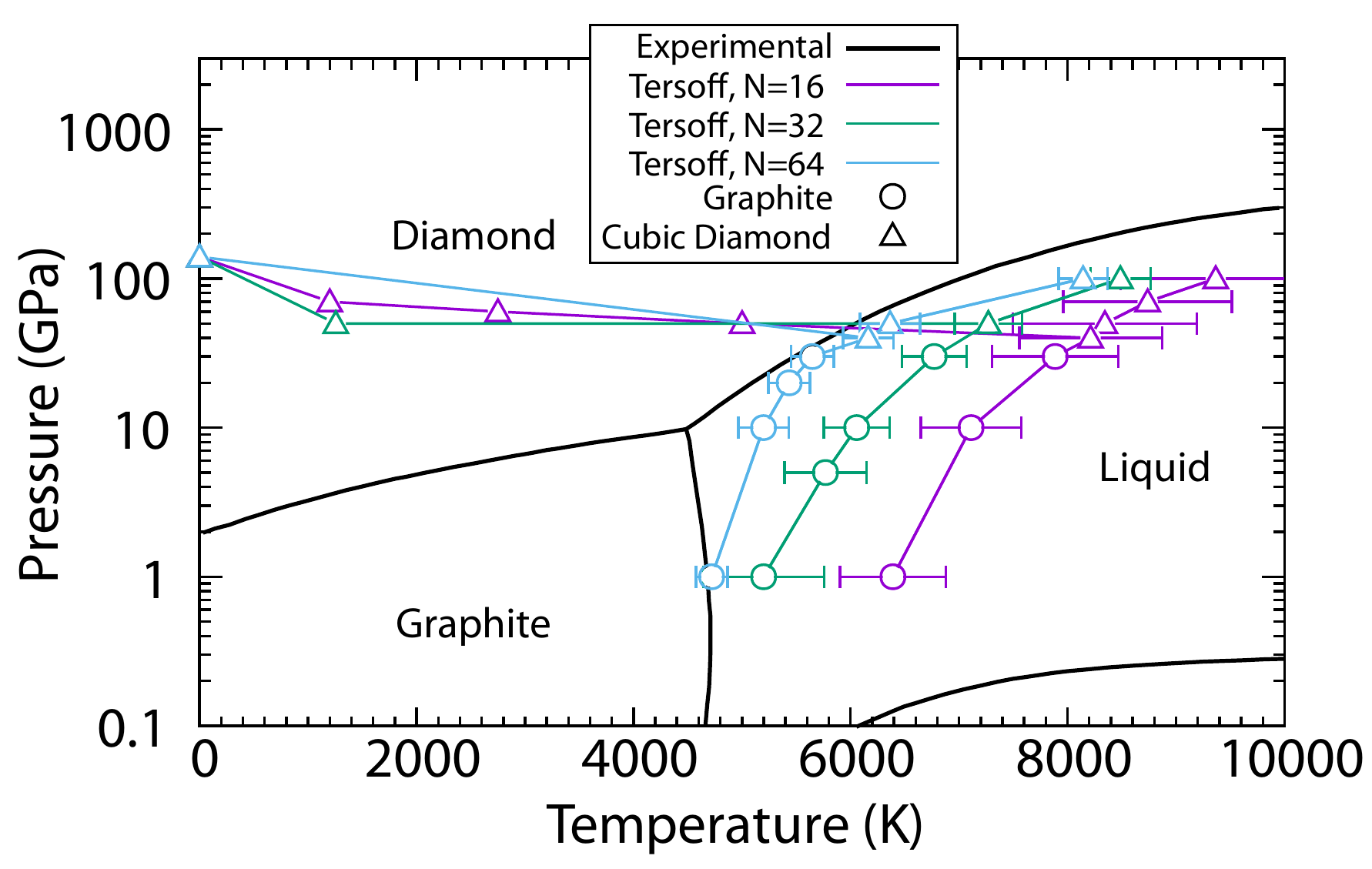}
\end{center}
\caption {Temperature-pressure phase diagram of the Tersoff potential. Black lines show experimental phase boundaries,\cite{bundy1989pressure,bundy1996pressure, steinbeck1985model} coloured lines and symbols correspond to nested sampling results with different system sizes. Error bars represent the full widths at half maximum of the heat capacity peaks.}
\label{fig:Tersoff_PD}
\end{figure}

\begin{figure}[hbt]
\begin{center}
\includegraphics[width=8.5cm,angle=0]{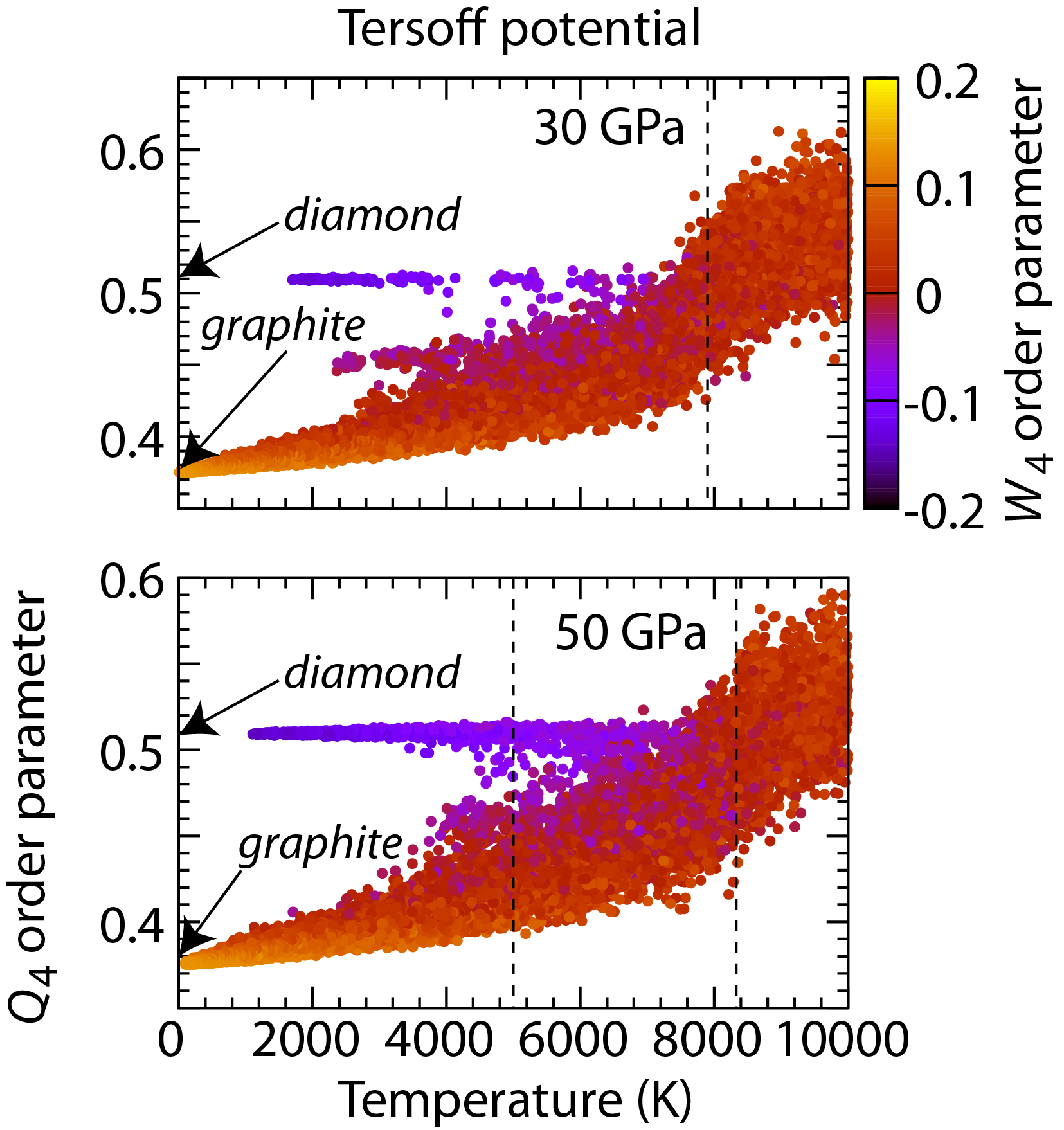}
\end{center}
\caption {Average $Q_4$ bond order parameter\cite{bib:Q6parameter} of configurations, generated by nested sampling, using the Tersoff potential at two different pressures. Each point corresponds to a configuration generated by nested sampling and coloured according to the average $W_4$ order parameter. Arrows point to the $Q_4$ values of diamond and graphite structures. Vertical dashed lines represent the phase transitions as determined by the peaks of the heat capacity curves. 
}
\label{fig:Tersoff_QW}
\end{figure}

The phase diagram calculated with the Tersoff potential is shown in Figure~\ref{fig:Tersoff_PD}.
Compared to the GAP-20 model, the Tersoff potential shows a significantly larger finite-size effect, which remains consistent at higher pressures as well. 
Overall the melting line reflects the experimental trend reasonably well at 64 atoms, however, the corresponding solid phases are different. Below 50~GPa, graphite is formed, but the melting line does not reflect the expected negative gradient at lower pressures, nor the significant change in the melting line gradient above the graphite-diamond-liquid triple point, which is overestimated compared to experimental results. The origin of Tersoff's monotonic melting curve in the graphite phase is its small cutoff of 4.1~\AA, which leads to a dramatic underestimation of the equilibrium lattice spacing compared to DFT, by around 40\%. This corresponds to a graphite phase that is more dense than the liquid phase at all pressures, hence the lack of a maximum in the melting curve.

Using the Steinhardt bond-order parameters\cite{bib:Q6parameter} $Q_4$ and $W_4$ we are able to distinguish diamond and graphite configurations generated by nested sampling, and locate the phase transition between different crystalline structures. 
In Figure~\ref{fig:Tersoff_QW} we demonstrate this at three different pressures. At 30~GPa, the large majority of the solid configurations fall into the basin of the graphite structure, however, the metastable diamond phase is also sampled to a lesser extent. A third and smaller basin (appearing to have $Q_4=0.45$) can be observed between these, representing a structure where small graphite-like motifs are interconnected by four-coordinated carbon atoms. As the pressure increases, the diamond structure becomes more dominant, until it becomes the ground state  structure at 80~GPa.
Due to the Tersoff potential's short range, the potential energy of the perfect cubic and hexagonal diamond structures are the same, and at pressures where the diamond phases are stable their sampling is about equal, suggesting that their free energy is comparable as well.


\subsection{EDIP}

\begin{figure}[hbt]
\begin{center}
\includegraphics[width=8.5cm,angle=0]{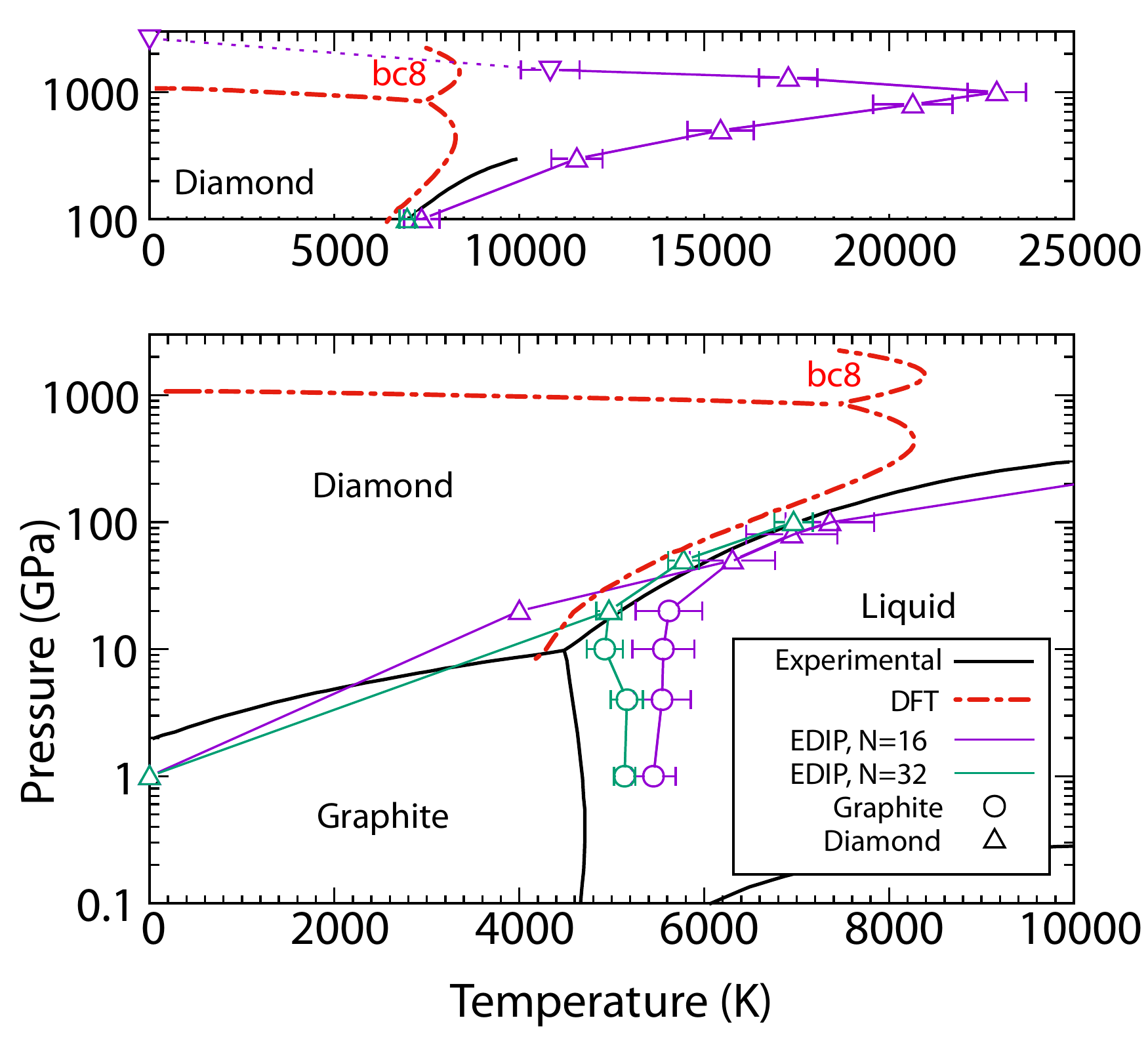}
\end{center}
\vspace{-20pt}
\caption{Temperature-pressure phase diagram of the EDIP potential. Black lines show experimental phase boundaries\cite{bundy1989pressure,bundy1996pressure, steinbeck1985model}, red dashed lines show high-pressure phase transitions predicted by DFT from Ref.\cite{C_DFT_highP_diagram} (with the phase above 1~TPa being bc8), purple and green coloured lines and symbols show nested sampling results with different system sizes. Error bars represent the full widths at half maximum of the heat capacity peaks. The top panel includes the high-pressure range of the phase diagram at a different temperature scale to show the maximum of the melting line.}
\label{fig:EDIP_PD}
\end{figure}

\begin{figure}[hbt]
\begin{center}
\includegraphics[width=8.5cm,angle=0]{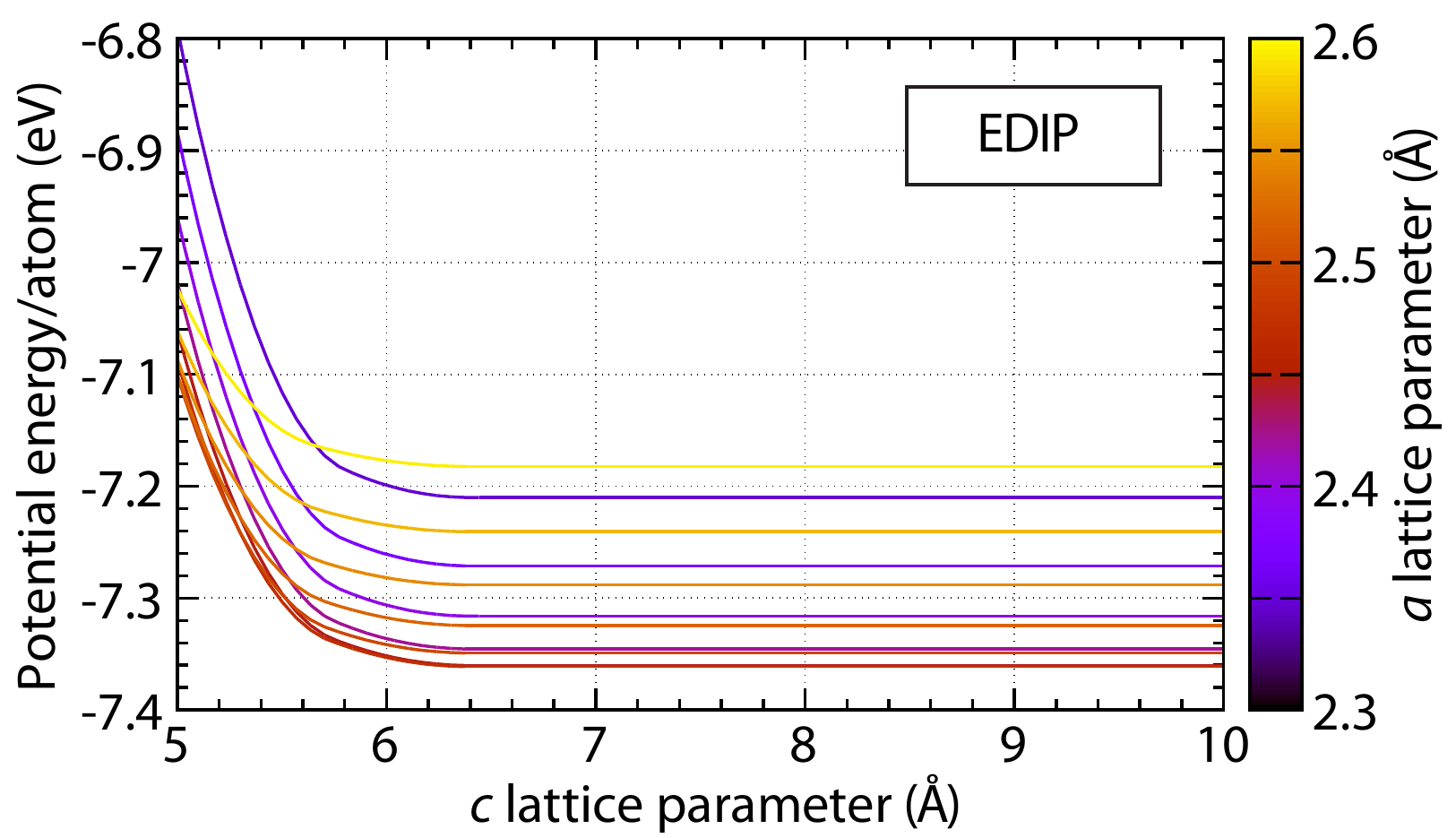}
\end{center}
\caption {Potential energy of the graphite AB structure as a function of lattice parameters, calculated using the EDIP potential.}
\label{fig:GAP_graphite-lattice_edip}
\end{figure}

\begin{figure}[hbt]
\begin{center}
\includegraphics[width=10.5cm,angle=90]{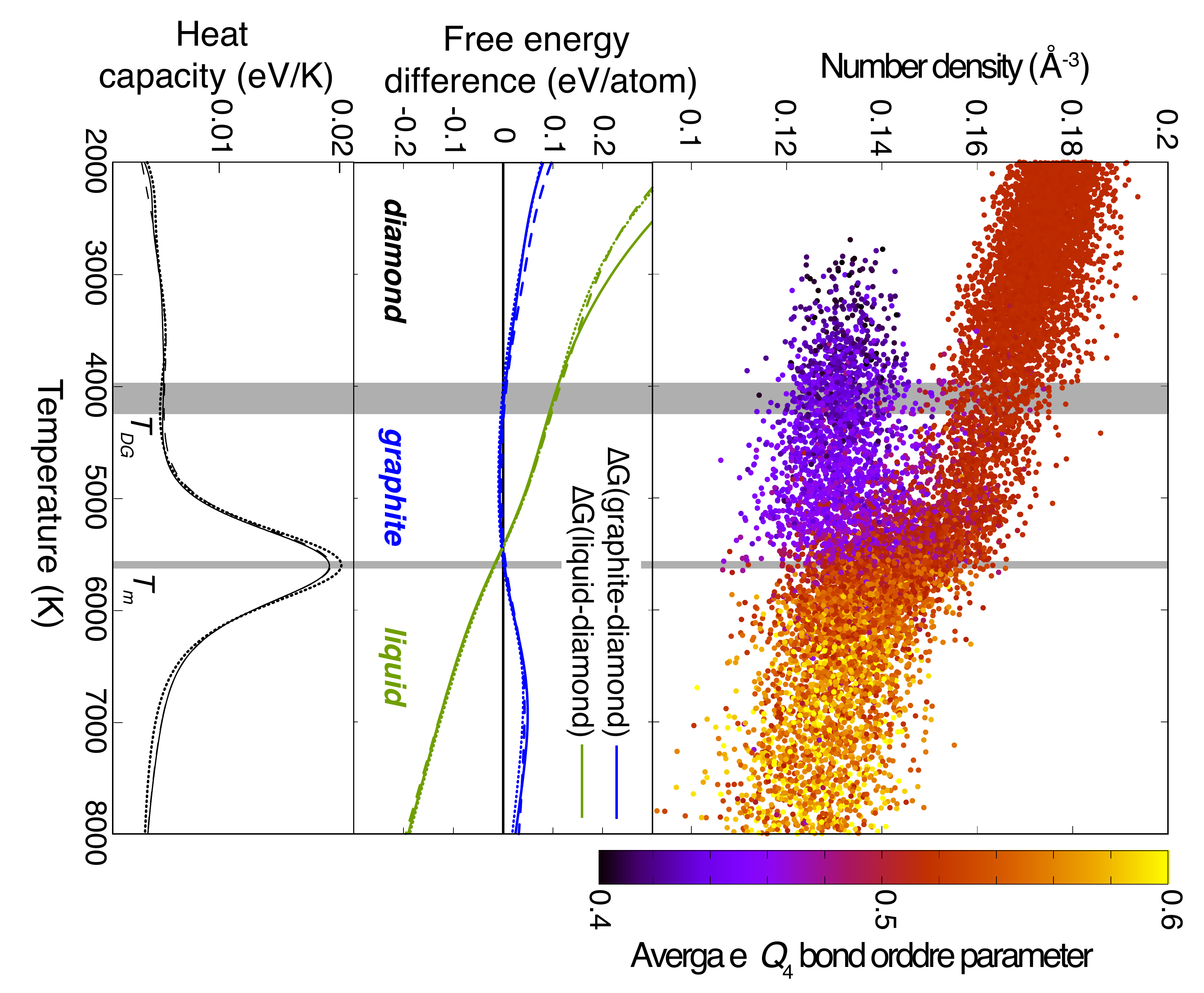}
\end{center}
\caption{Nested sampling results at 20~GPa, using the 16 atoms and the EDIP potential. Top panel: number density of individual configurations sampled during NS, symbols are coloured by the average $Q_4$ bond order parameter of the configuration. Middle panel: Gibbs free-energy difference compared to the diamond phase in three parallel runs. [Configurations were associated with basins, using the number density and the bond order parameter.] Bottom panel: Heat capacity curves of the three parallel runs.
Vertical grey lines show the melting temperature, $T_m$ and the solid-solid transition between the diamond and graphite phases, $T_{DG}$.}
\label{fig:EDIP_QW}
\end{figure}

\begin{figure}[hbt]
\begin{center}
\includegraphics[width=8cm]{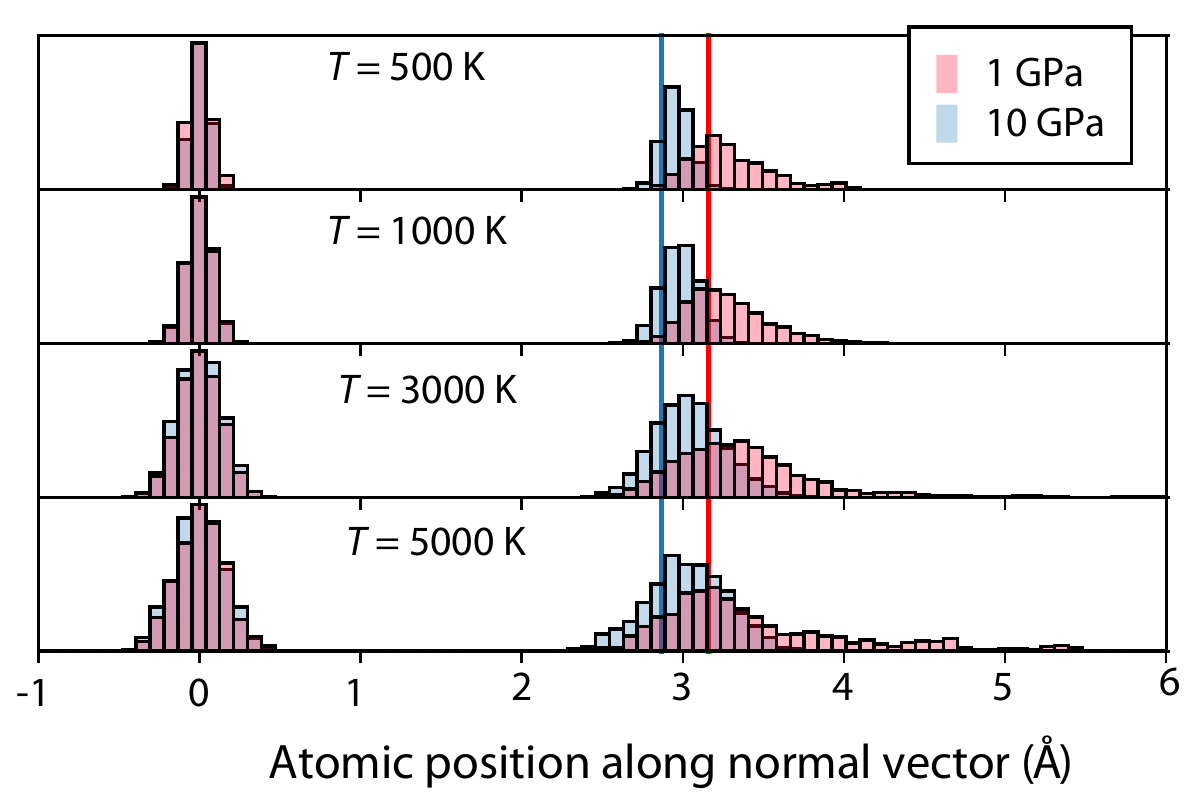}
\end{center}
\vspace{-20pt}
\caption {Distribution of carbon atoms along the direction perpendicular to the graphite layers, calculated as the weighted average from NS configurations, using the EDIP potential. Distribution around 0.0~\AA~acts as the reference layer, its widths representing the deviation from a perfectly flat layer. Bars are semi-transparent to aid the visibility of the distributions. Vertical lines show the equilibrium graphite layer distance at the two different pressures.}
\label{fig:EDIP_layers}
\end{figure}

Nested sampling calculations using the EDIP potential were performed with 16 and 32 atoms, at pressures ranging from 1 GPa to 1500 GPa. 
The resulting phase diagram is shown in Figure~\ref{fig:EDIP_PD}, showing an overall excellent agreement with experimental phase behaviour up to 100 GPa. The melting line follows the overall trend well, with a considerably smaller finite size effect compared to the Tersoff potential. 
At lower pressures graphite is formed upon freezing, as expected, with typical layer spacings at low temperature corresponding to a lattice parameter of $c=6.4$~\AA, only a 5\% underestimation of experimental data. To explore the EDIP's graphite phase further, we plot its energy as a function of lattice parameters in Figure~\ref{fig:GAP_graphite-lattice_edip}. One of the potential's shortcomings is its lack of dispersive, long-range interactions, and that is reflected in its graphitic energy landscape, as we see no change in energy beyond $c=6.4$~\AA. While this is not consistent with the clearly defined minimum separation predicted by DFT, the influence of the $PV$ term in the enthalpy effectively prevents larger separations from being energetically relevant at finite pressures and zero temperature. To evaluate the finite temperature effect of this short interplanar cutoff, we plot thermally averaged distributions of carbon atoms perpendicular to the graphite planes in Figure~\ref{fig:EDIP_layers}, for pressures of 1, 4 and 10 GPa. These show an expected broadening of carbon atom dispersion in the reference layer at higher temperatures, due to thermal disorder, but for the nearest-layer distributions this broadening becomes more biased towards larger spacings as temperature increases, suggesting that the lack of a long-range energy barrier allows un-physically large layer separations to overcome the $PV$ term and become thermodynamically relevant.

This large separation-bias persists at higher pressures, however the effect is diminished, which can be intuitively understood as the increased pressure (and $PV$ energy) encouraging smaller volumes, and preventing thermal fluctuations from stabilising larger-separation structures. As the temperature decreases, there is less kinetic energy available to smear the energies of the optimised structure, and thus fewer large separation configurations can be energetically viable. In spite of EDIP's short interplanar cutoff and its effects, its description of the graphite melting line is remarkably accurate at only 32 atoms. Given the importance of the ratio between liquid and solid densities in shaping the melting line, and that graphite's volume is particularly sensitive to interplanar separation, these results show that an accurate description of graphite's phase space is essential for determining its melting behaviour.

At a pressure of 10 GPa we begin to observe a small number of cubic and hexagonal diamond structures among the sampled configurations, around the freezing transition. However, these structures quickly lose thermodynamic relevance in comparison to the graphite phase. When pressure is increased to 20 GPa, the diamond configurations become energetically viable enough that the NS algorithm simultaneously samples them along with the graphite phase, such that we can identify a solid-solid transition. 
Figure~\ref{fig:EDIP_QW} shows the configurations sampled by NS, and using appropriate order parameters, we are able to sort these into different structural basins. This allows us to calculate the contribution of each basin (structure) to the partition function separately, and hence their free energy. The free energy of the diamond and graphite structures are shown in the middle panel of Figure~\ref{fig:EDIP_QW}, demonstrating that below the melting point graphite is more stable than diamond, however their free energy difference is very small. 
Like in the case of the Tersoff potential, the potential energy of the perfect cubic and hexagonal diamond structures are the same using EDIP, and above 20~GPa NS runs sampled both structures equally, suggesting that their free energy is comparable.


\section{Conclusion}

In the current work we reviewed the performance of three interatomic potential models of carbon, focusing on their ability to reproduce experimentally observed macroscopic properties.
We used the nested sampling technique to sample the potential energy surface of these models in a wide pressure range, calculating their pressure-temperature phase diagram and predicting crystalline phases.
We emphasise that nested sampling is a unique tool that allows us the exhaustive exploration of the phase space and makes the calculation of the entire phase diagram a relatively straightforward process, while also being predictive and not restricted by known or considered crystalline structures.
All three models, GAP-20, Tersoff and EDIP, predicted the graphite structure to be more stable at low pressures and the diamond structure at higher pressures, however, the transition between these as well as the location of the melting line differed considerably. 
Empirical potentials are often fitted to specific microscopic properties, for example to typical coordination of graphite and diamond structures, hence their high-temperature and high-pressure behaviour cannot be expected to reflect accurately the diverse structural properties of carbon. 
Nevertheless, while the macroscopic properties of the Tersoff potential differ from the experimental phase diagram considerably, we found that the phase diagram of the EDIP potential to be very close to experimentally observed behaviour, both the predicted graphite-diamond transition, as well as the melting line up to relatively high pressures.

Machine learning (ML) potentials provide the state-of-the-art description of atomic interactions, opening up new routes to materials discovery that are otherwise out of our reach, offering \textit{ab initio} level accuracy at an affordable computational cost. 
However, the main criticism of ML potentials is that they are inherently best suited only to interpolation problems, and perform reliably only in the regime of configuration space where the potential was trained. 
This means that their behaviour in unexplored territory, in the region where the potential is forced to use extrapolation, can be unrealistic or unphysical, limiting or even inhibiting their use in scientific discovery. 
Therefore, our results showing that the GAP-20 potential performs well and predicts the expected phase transitions reliably up to 300 GPa, well outside the original training conditions, is remarkable, emphasising the power of including a diverse range of atomic local environments in the training process. 
Moreover, the exhaustive exploration provided by NS also highlighted local weaknesses of the model, such as the unexpected graphite-layer distances or the very high pressure phases, offering areas for potential improvement and extensions of the GAP-20 model.

\begin{acknowledgements}
The authors thank Miguel Caro for providing DFT parameters consistent with training of the GAP-20u potential, and Nigel Marks for providing access to the carbon EDIP. The authors also thank Albert P. Bart\`ok, G\`abor Cs\`anyi and Volker Deringer for useful discussions around the performance of the GAP-20 model.
L.B.P. and B.K. acknowledge support from the EPSRC through the individual Early Career Fellowships (LBP: EP/T000163/1 and BK: EP/T026138/1).
Computing facilities were provided by the Scientific Computing Research Technology Platform of the University of Warwick. 
Calculations using the GAP potential were performed using the Sulis Tier 2 HPC platform hosted by the
Scientific Computing Research Technology Platform at the University of Warwick.
Sulis is funded by EPSRC Grant EP/T022108/1 and the HPC Midlands+ consortium.
\end{acknowledgements}



\bibliography{carbon_references}

\end{document}